\documentclass[%
reprint,
letterpaper,
10pt,
article,
superscriptaddress,
twocolumn,
nofootinbib,
amsmath,amssymb,
aps,
showkeys,
prl,
floatfix,
]{revtex4-2}
\usepackage{dcolumn}% Align table columns on decimal point
\usepackage{bm}% bold math
\usepackage[hyperindex=true,colorlinks=true,linkcolor=blue,citecolor=blue,urlcolor=blue]{hyperref}
\usepackage{xcolor}
\DeclareUnicodeCharacter{0300}{e}
\usepackage[T1]{fontenc}
\usepackage{graphicx}
\usepackage{epstopdf, epsfig}
\usepackage[utf8]{inputenc}
\usepackage{upgreek}
\usepackage{relsize}
\usepackage[normalem]{ulem}
\usepackage{footmisc}
\usepackage{braket}
\usepackage{ulem}
\usepackage{enumitem}
\usepackage{gensymb}

\setlist{nosep}
\allowdisplaybreaks[1]

\begin{document}

\title{Experimental evidence for coronal mass ejection suppression in strong stellar magnetic fields}

\author{S.N. Chen}
\affiliation{ELI NP, ”Horia Hulubei” National Institute for Physics and Nuclear Engineering, 30 Reactorului Street, RO-077125, Bucharest-Magurele, Romania}
\author{K. Burdonov}
\affiliation{LULI - CNRS, CEA, Sorbonne Université, Ecole Polytechnique, Institut Polytechnique de Paris - F-91128 Palaiseau cedex, France}
\affiliation{Sorbonne Université, Observatoire de Paris, Université PSL, CNRS, LUX, F-75005, Paris, France}
%\affiliation{IAP, Russian Academy of Sciences, 603155, Nizhny Novgorod, Russia}

\author{W. Yao}
\affiliation{LULI - CNRS, CEA, Sorbonne Université, Ecole Polytechnique, Institut Polytechnique de Paris - F-91128 Palaiseau cedex, France}
\affiliation{Sorbonne Université, Observatoire de Paris, Université PSL, CNRS, LUX, F-75005, Paris, France}

\author{J.~D. Alvarado-Gómez}
\affiliation{Leibniz Institute for Astrophysics Potsdam, An der Sternwarte 16, 14482 Potsdam, Germany}

\author{C. Argiroffi}
\affiliation{INAF, Osservatorio Astronomico di Palermo, Palermo, Italy}
\affiliation{University of Palermo, Department of Physics and Chemistry, Palermo, Italy}

\author{J. Béard}
\affiliation{LNCMI, UPR 3228, CNRS-UGA-UPS-INSA, 31400, Toulouse, France}

\author{S. Bolanõs}
\affiliation{LULI - CNRS, CEA, Sorbonne Université, Ecole Polytechnique, Institut Polytechnique de Paris - F-91128 Palaiseau cedex, France}

\author{R. Bonito}
\affiliation{INAF, Osservatorio Astronomico di Palermo, Palermo, Italy}

\author{A. Ciardi}
\affiliation{Sorbonne Université, Observatoire de Paris, Université PSL, CNRS, LUX, F-75005, Paris, France}
\author{O. Cohen}
\affiliation{Lowell Center for Space Science and Technology, University of Massachusetts Lowell, 600 Suffolk St., Lowell, MA 01854, USA}

\author{J.~J.~Drake}
\affiliation{Center for Astrophysics $|$ Harvard \& Smithsonian, 60 Garden Street, Cambridge, MA 02138, USA}
\affiliation{Lockheed Martin Solar and Astrophysics Laboratory, 3251 Hanover Street, Palo Alto, CA 94306, USA}

\author{S. Orlando}
\affiliation{INAF, Osservatorio Astronomico di Palermo, Palermo, Italy}
\author{J. Fuchs}
\email{julien.fuchs@polytechnique.edu; julien.fuchs@technion.ac.il}
\affiliation{LULI - CNRS, CEA, Sorbonne Université, Ecole Polytechnique, Institut Polytechnique de Paris - F-91128 Palaiseau cedex, France}
\affiliation{Technion Israel Institute of Technology, Faculty of Physics, Technion City, Haifa 3200003, Israel}

\date{\today}

\begin{abstract}
Solar coronal mass ejections (CME) are routinely observed, but as of yet there exist few convincing detections of stellar CMEs. A reason for this could be the stronger magnetic fields of these stars, compared to that of our Sun, would prevent CME to form and escape. Here we combined astrophysical simulations, measurements of scaled high-energy laser–driven plasma flows, and 3D magneto-hydrodynamic modeling to test this hypothesis. %We found that, depending on the ratio of the plasma pressure versus the magnetic pressure, there are three behaviors: unimpeded propagation, partial confinement, or full halting of the flow. When the applied laboratory magnetic field is $10^5$ G, i.e. scaling equivalent to 30 G of a stellar object, the plasma stream can propagate unimpeded. However, when the magnetic field is increased to $3 \times 10^5$ G, which scales to a stellar 100 G field, the stream is observed to stop or even recede.
Simulations show that in a 100 G stellar dipole field, low–plasma $\beta$ CMEs become magnetically confined. In the laboratory, a laser-produced plasma stream scaled to stellar CME conditions propagates freely at low applied magnetic fields (approximately 30 G stellar equivalent) but becomes unstable and halts entirely when the field is increased to 3e5 G (i.e., a ~100 G equivalent). Numerical simulations suggest that the sudden disruption of the flow is induced by a kink instability. These results provide the first laboratory-scale evidence that strong stellar magnetic fields can fully suppress CME propagation, offering a physical explanation for their lack in stellar observations and highlighting the role of magnetic confinement in stellar evolution and exoplanet space weather.%These experimental data and numerical simulations thus shed light on the dichotomy of CME behavior between our Sun and active stars. % provide  evidence that stellar CMEs can be suppressed by magnetic activity of a star.
\end{abstract}

%\pacs{Valid PACS appear here}
% PACS, the Physics and Astronomy Classification Scheme.

\maketitle

\section{Introduction}

Coronal mass ejections (CMEs) are large-scale expulsions of plasma and magnetic field from stellar atmospheres. On the Sun, they are routinely observed and play a central role in mass and angular momentum loss, as well as in shaping the near-planetary environment \cite{Aarnider_2012,Drake_2013,Osten_2015,10.1093/mnras/stx1969}. In the context of exoplanetary systems, CMEs from host stars can drive intense space weather, with profound implications for atmospheric erosion and habitability \cite{KHODACHENKO2007631,doi:10.1089/ast.2006.0127,doi:10.1089/ast.2006.0128,airapetian_2020}.

Despite their importance, direct detections of CMEs from stars other than the Sun remain rare. Decades of observational efforts, ranging from H$\alpha$ spectroscopy \cite{Vidaetal19} to searches for X-ray coronal dimmings \cite{Veronig2021} and Type II radio bursts \cite{Crosley_2018}, have yielded only a handful of candidates, and in most cases the inferred kinetic energies are far below expectations from solar flare-CME scaling laws \cite{Moschou_2019,Argiroffi2019,Namekata2022}. This discrepancy has emerged as one of the central puzzles in stellar activity research.

In stars, the magnetic field correlates inversely with the star's rotation period through magnetic dynamo action. The magnetic field at the stellar surface ranges from 10 - 1000 G, as inferred from  astronomical observations \cite{doi:10.1146/annurev-astro-082708-101833,2014MNRAS.441.2361V}%, hence giving most active stars kilo-Gauss scale surface magnetic fields \cite{Saar.90,10.1093/mnras/stu728,10.1093/mnras/stv2924,10.1093/mnras/stx3021}
. Therefore, it has been speculated that these fields, which are much higher than that of our Sun, could produce a significant countering effect on CME propagation \cite{10.1093/mnras/stx1969,Alvarado_G_mez_2018,Alvarado_G_mez_2019}. Theoretical investigations of the behavior of stellar CMEs with magnetic fields stronger than the solar field have been made recently \cite{Alvarado_G_mez_2018,Alvarado_G_mez_2019}. It was found that the strong magnetic field causes a substantial reduction of the radial velocity of the eruptions, and if the field strength is increased, it is able to fully confine the simulated CME. This may explain why the few detected candidate CMEs in active stars are characterized by energies much lower than those expected from the extrapolation of the solar flare-CME relation to large flare energies. Furthermore, while theoretical studies provide a plausible mechanism, direct experimental evidence for CME suppression under astrophysically relevant conditions has been lacking. Bridging this gap requires an approach that can probe the plasma-magnetic field interaction in a controlled, scalable setting.

Here we address this problem by combining three complementary methods. First, we use the Alfven Wave Solar Model (AWSoM-R) to simulate the evolution of a CME-like perturbation in a stellar corona, to see if our scaled laboratory conditions could possibly reproduce plasma confinement. Second, we design and perform high-energy laser–plasma experiments in which a supersonic, magnetized plasma flow propagates across a transverse magnetic field. \textcolor{black}{The expanding plasma in the experiment is designed to model the front of a CME interacting with a transverse magnetic field, after it has detached from the hot flare loop at the stellar surface \cite{Chen2017,Lin2015}. This complements experiments performed with discharge plasmas aimed at modeling the early dynamics of arches, i.e. when they are located close to the surface \cite{Hansen2001,Sklodowski2023}. We also note that the configuration we model is different than that of tokamaks plasmas, since there is here a strong outward ram pressure, associated to the CME launching in the astrophysical case, and to the laser-associated ablation pressure in the laboratory experiment. The laboratory plasma is }scaled to stellar CME conditions using the Euler similarity criteria \textcolor{black}{established by Ryutov et al. \cite{Ryutov_1999,Ryutov_2000}}. By varying the applied field strength, we directly observe the transition from unimpeded propagation to complete flow halting. Finally, we perform 3D resistive MHD simulations with the
code to investigate the instability dynamics responsible for disruption, identifying a kink mode as the dominant mechanism at high magnetic field strengths \cite{yao2022characterization}.

\begin{table*}
\centering{}%
\begin{tabular}{cc||c|c|c}
 & \multicolumn{1}{c||}{\textbf{\textit{Stellar CME}}} & \multicolumn{1}{c}{\textbf{\textit{}}}  & \multicolumn{1}{c}{\textbf{\textit{Laboratory}}} & \multicolumn{1}{c}{\textbf{\textit{}}}\tabularnewline

\hline
\hline
Material & \multicolumn{1}{c||}{$H$} & \multicolumn{1}{c}{} & \multicolumn{1}{c}{$CF_{2}$ (teflon) } & \multicolumn{1}{c}{}\tabularnewline

Charge state Z & \multicolumn{1}{c||}{$\textbf{1}$} & \multicolumn{1}{c}{} & \multicolumn{1}{c}{$\textbf{5.9}$} & \multicolumn{1}{c}{}\tabularnewline

Mass number A & \multicolumn{1}{c||}{$\textbf{1.28}$} & \multicolumn{1}{c}{} & \multicolumn{1}{c}{$\textbf{17.3}$} & \multicolumn{1}{c}{}\tabularnewline

Spatial scale $L$ $[cm]$ & \multicolumn{1}{c||}{$\mathbf{2 \times 10^{11}}$} & \multicolumn{1}{c}{} & \multicolumn{1}{c}{$\mathbf{1}$} & \multicolumn{1}{c}{}\tabularnewline

%spatial scale $[cm]$ & $\mathbf{0.1}$ & $\mathbf{}$ & $\mathbf{5 \times 10^{11}}$ & $\mathbf{7 \times 10^{11}}$\tabularnewline

Magnetic field $B$ $[G]$ & \multicolumn{1}{c||}{$\mathbf{100}$} & \multicolumn{1}{c}{} & \multicolumn{1}{c}{$\mathbf{3 \times 10^{5}}$} & \multicolumn{1}{c}{}\tabularnewline

\textcolor{black}{Time Scale} & \multicolumn{1}{c||}{\textcolor{black}{60 minutes}} & \multicolumn{1}{c}{} & \multicolumn{1}{c}{\textcolor{black}{10 ns}} & \multicolumn{1}{c}{}\tabularnewline

\hline

Alfven Mach number $M_{Alf}$  & \multicolumn{1}{c||}{7.6} & \multicolumn{1}{c}{} & \multicolumn{1}{c}{4.3} & \multicolumn{1}{c}{}\tabularnewline

 Reynolds number $R_{e}$ & \multicolumn{1}{c||}{$1.5 \times 10^{11}$} & \multicolumn{1}{c}{} & \multicolumn{1}{c}{$1.3 \times 10^{4}$} & \multicolumn{1}{c}{}\tabularnewline

Magnetic Reynolds number $R_{eM}$ & \multicolumn{1}{c||}{$3.2 \times 10^{18}$} & \multicolumn{1}{c}{} & \multicolumn{1}{c}{$5 \times 10^{4}$} & \multicolumn{1}{c}{}\tabularnewline

Peclet number $P_e$ & \multicolumn{1}{c||}{$3.1\times 10^{9}$} & \multicolumn{1}{c}{} & \multicolumn{1}{c}{$74$} & \multicolumn{1}{c}{}\tabularnewline

Euler number $Eu$  & \multicolumn{1}{c||}{22} & \multicolumn{1}{c}{} & \multicolumn{1}{c}{15} & \multicolumn{1}{c}{}\tabularnewline

Thermal plasma $\beta_T$ & \multicolumn{1}{c||}{$0.24$} & \multicolumn{1}{c}{} & \multicolumn{1}{c}{$0.17$} & \multicolumn{1}{c}{}\tabularnewline

\textcolor{black}{Ion Larmor Radius / Spatial scale  ($r_{Li}/L$)}  & \multicolumn{1}{c||}{\textcolor{black}{$1\times 10^{-10}$}} & \multicolumn{1}{c}{} & \multicolumn{1}{c}{\textcolor{black}{$1.6\times 10^{-3}$}} & \multicolumn{1}{c}{}\tabularnewline

\tabularnewline
\end{tabular}
\caption{Comparison and scalability between a hypothetical stellar CME event, propagating across a magnetic field of 100 G and the laboratory plasma stream, created by the nominal laser irradiance, that is disrupted during its propagation across the magnetic field of 3$\times10^5$ G. }
\label{tab.scal_2R}
\end{table*}

\section{Astrophysical Simulations}
\label{sec:astro.sim}

%Now
We first %To test if the plasma regime should exhibit a %the expected
%suppression behavior, we
simulated the evolution of a local perturbation representing the core of a stellar CME, and determined the plasma conditions under which %whether or not
it is or is not able to escape the magnetic confinement imposed by the stellar magnetic field.

With regard to plasma conditions of stellar CMEs, we have limited observational data and there is almost no information about typical CME parameters \cite{Moschou_2019}. One exception is that presented in \citealt{Argiroffi2019} which, however, was a very extreme event. Note that we do not consider here detected flare properties to correspond to CME plasma parameters but rather the properties of prominence eruptions \cite{Namekata2021,Leitzinger2024}.
Some estimations of plasma conditions can be found in \citealt{10.1093/mnras/stx1969} and \citealt{10.1093/mnras/staa1021}, supposing CME velocities from hundreds to thousands of kilometers per second, and effective star surface plasma temperatures of the order of hundreds of kKelvins (while the CME itself is characterized by temperature in the MKelvins range).

With these estimates, we can find a set of conditions of a hypothetical CME that relates well to our laboratory plasmas. To do this, and in order to investigate in the laboratory the astrophysical plasmas in a relevant manner, we use the Euler similarity approach developed in the works by Ryutov \cite{Ryutov_1999,Ryutov_2000,doi:10.1063/1.5042254} (for more scaling details, see the Supplementary Information). The idea here is that the two systems will evolve in a similar fashion, when the two scaling quantities, the Euler number and the plasma  $\beta$, are the same for the two systems. \textcolor{black} {Here the Euler number is defined as $ E_u = v \sqrt{\rho/p}$  where $v$ is fluid velocity, $\rho$ is the plasma density, and $p$ is the plasma pressure; it physically represents the balance between the ram and plasma pressure.  The plasma  $\beta$  is the ratio of the magnetic pressure and plasma pressure. This approach is valid if the two systems can be considered ideal MHD with the addition of the following conditions: (1) the hydrodynamic (convective) transport needs to be dominant, hence the Peclet number, which is the ratio of heat convection to heat conduction, needs to fulfill $P_e$  >> 1; (2) the viscosity is small, hence the Reynolds number $R_e$ >> 1; and (3) the ohmic dissipation is negligible i.e. high electrical conductivity, hence the magnetic Reynolds number $R_{eM}$ >> 1.}

\textcolor{black} {The parameters of a possible stellar CME, and those of the laboratory plasma are  detailed in Table 1 and in the Supplementary Information. We can see that both configurations are indeed in the ideal MHD regime, but that they further share similar Alfen Mach numbers. The similarities between the Euler and plasma  $\beta$ numbers of both systems ensures that they are scalable to each other. We furthermore see that both are strongly magnetized, since they both have $r_{Li}/L$ << 1. Finally, since for both we have $r_{Li}/L$ << $(\beta_T)^{3/2}$, where $\beta_T$ is the ratio of the magnetic pressure and plasma thermal pressure, this also ensures that non-MHD physics, associated with  Hall and kinetic terms,  can be neglected \cite{Khiar2017}.}

%An abridged version of our calculations is shown in Table 1, with more details in the Supplementary Information.

\begin{figure}
\includegraphics[width=0.48\textwidth]{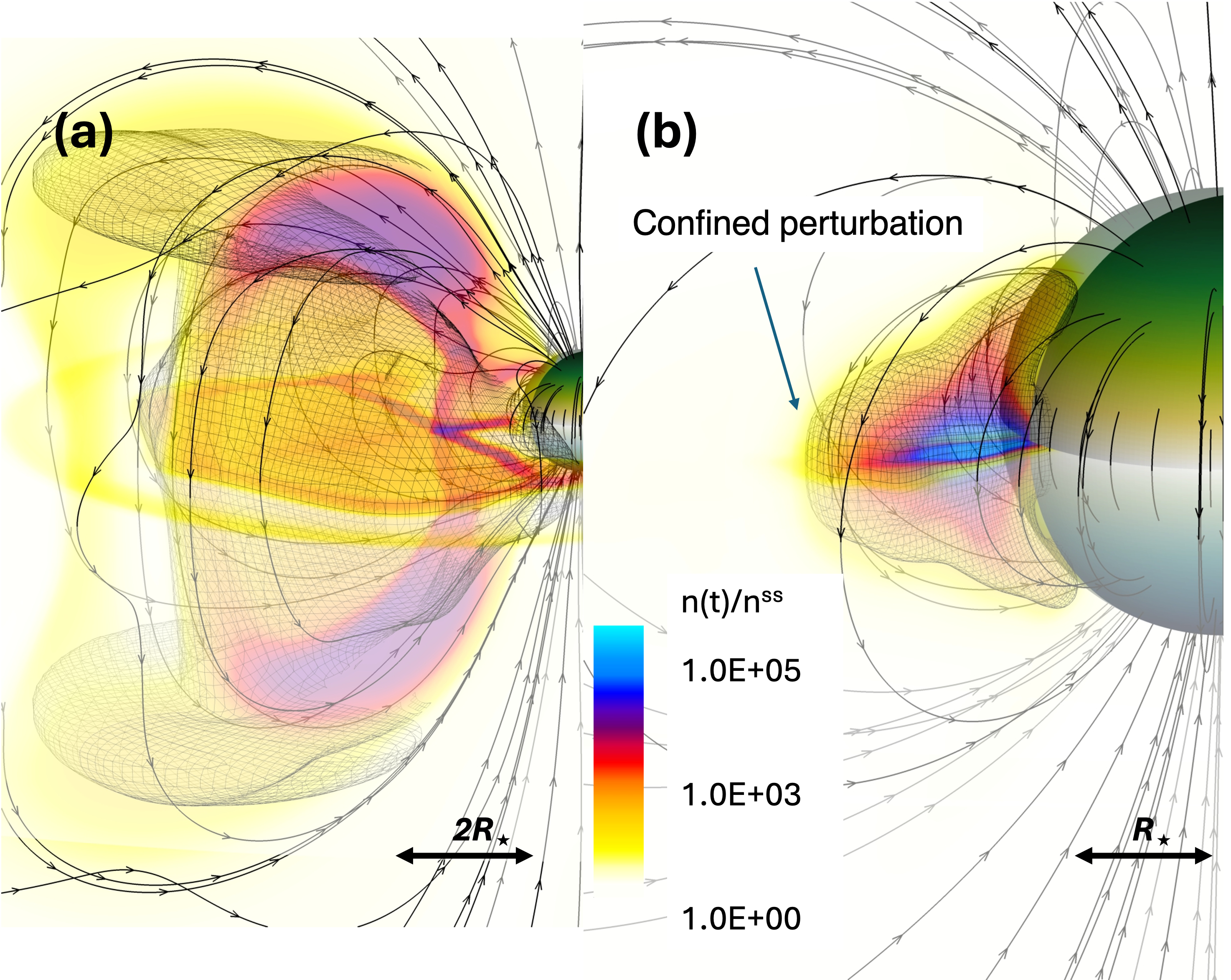}
\caption{Simulation results, using the AWSoM-R code, of the evolution of a hypothetical stellar CME event to which our laboratory plasmas (see Table 1) are scaled to. Snapshots during the temporal evolution of a spherical perturbation at t=60 min with a initial temperature of (a)  $T_{\rm p} = 5 \times 10^{7}$~K and (b)  $T_{\rm p} =  10^{7}$~K. We are able to see the perturbation plasma is confined within the stellar magnetic fields, with the lower temperature plasma being more susceptible to the magnetic fields. \textcolor{black}{The quantity $n(t)/n^{SS}$ shown in the colorbar corresponds to the instantaneous density contrast between the particle density at time ‘t’ and  the background steady-state density (taken at t=0.0). In standard solar/stellar CME simulations, the expanding CME front is identified by the ratio $n(t)/n^{SS}$ = 3.0 and the core of the eruption displays values $n(t)/n^{SS} \ge$ 10.0. Both simulations are showing the snapshot at t = 60 min with the gridded iso-surface denoting the $n(t)/n^{SS}$ = 10.0 region.}}
\label{fig:astro_simu}
\end{figure}

To simulate the stellar CMEs, we employed the faster-than-real-time Alfv\'en Wave Solar Model (AWSoM-R \cite{2014ApJ...782...81V, 2021ApJ...908..172S}), which is part of the open-source Space Weather Modelling Framework (SWMF) to simulate the evolution of a local perturbation, representing the core of a CME. To represent the star, a simple $100~$G dipole aligned with the stellar rotation axis %, taken in the direction of the cartesian $z-$axis
was used as the large-scale magnetic field of the star. \textcolor{black}{This choice of  magnetic field strength and idealised geometry for the astrophysical simulation was motivated by the set-up implemented in the laboratory experiments, but is also  comparable to observational constraints obtained via spectropolarimetry and Zeeman-Doppler Imaging on the large-scale magnetic field structure of cool stars \cite{2022csss.confE.209D}. In particular, similar dipole field strengths and axisymmetric topologies have been reported on very young (~250-500 Myr old) Sun-like stars (see e.g. \cite{Folsom2017}) as well as in M-dwarfs of different ages (see \cite{Kochukhov2020}). Moreover, after initial expansion, the development of a CME depends mostly on the large-scale field that the dipolar configuration represents.}  We then consider a spherical perturbation ($R_{\rm p} = 0.15~R_{\bigstar}$) located at $R = 2R_{\bigstar}$, with particle density $n_{\rm p} = 10^{12}$~cm$^{-3}$ and temperature of $T_{\rm p} = 10^{7}$~K and  $5 \times 10^{7}$~K. \textcolor{black}{We underline that it is physically equivalent to vary the plasma pressure while keeping the magnetic field constant, and to maintain the plasma pressure but vary the magnetic field. We chose the first approach in order, similarly as in the laboratory experiment, to keep the same steady-state reference model over which the CME-like perturbation is inserted and tracked in time-accurate mode.} For our steady-state corona and wind model, we assume solar fiducial values for mass ($M_{\star} = M_{\odot}$) and radius ($R_{\star} = R_{\odot}$), as well as a 5-day rotation period consistent with the values observed in Sun-like stars displaying comparable large-scale magnetic field strengths. The corresponding thermal plasma $\beta_{T}$ values, with these initial conditions, are $\beta_{T} = 3.47$ and $\beta_{T} = 17.4$ respectively. The simulation is stopped at 60 minutes. \textcolor{black}{Note that this timescale might be insufficient to trace the propagation of a CME out to many stellar radii in a real astrophysical system (particularly up to distances comparable to the solar system planets). However, it is enough to test the CME suppression mechanism, i.e. whether or not the perturbation is confined by the large-scale magnetic field). Note moreover that} our scaling shows that a time period of 10 ns in our laboratory experiment (which is the time scale over which we observe flow disruption, see below) is scaled to 60 minutes in the astrophysical case. More details about the scaling can be found in the Supplementary Information.

The results of the simulations are shown in Fig.~1, showing the extent of the perturbation, in density, after 60 minutes.  A trend that can be observed: there is more confinement of the CME when  going to a lower initial temperature (compare Fig.~1.a and Fig.~1.b). This corresponds to decreasing the plasma $\beta $, since we are keeping all the other plasma parameters and magnetic field constant.

\section{The laboratory experiment}
\label{sec:lab.exp}

% \subsection{Experimental setup}

%The  laser-produced magnetized plasma flows that we are able to produce in the laboratory  experiment (see details below)~\cite{PhysRevLett.123.205001,Filippov2021}, shows that we can comfortably create plasmas with $ E_u = 15$ and  $\beta_T = 0.24$.

We conducted the experiment at the ELFIE laser facility (Laboratoire pour l'utilisation des lasers intenses, Ecole Polytechnique, France) \cite{Zou_2008}. The set-up we used is shown in Fig. 2a. A laser pulse (energy up to 50 J, duration 0.6 ns at FWHM, central wavelength 1057 nm) was focused onto the surface of a Teflon (CF$_{2}$) target, thus delivering an on-target intensity of $10^{13}$ W/cm$^{2}$.
\textcolor{black} {We  note that the target used is CF$_{2}$ and not the typical CME composition of hydrogen. In accordance with the scaling framework (see SI for the details) that we are using to relate our laboratory experiment to the astrophysical setting, the requirements of the values of the Peclet number, magnetic Reynolds number, and the Reynolds number, which have explicit dependence on the atomic mass and atomic number, are nonetheless satisfied with the plasma created using the CF$_{2}$ target material.}

\textcolor{black} {Following the laser ablation of the target material, a hot plasma flow is launched into the vacuum } %The main propagation of the induced hot plasma stream was
along the $z$-axis \textcolor{black} {and against the vacuum}. As shown in Fig. 2a, a large-scale quasi-static homogeneous magnetic field, with a strength up to $3 \times 10^5$ G using a Helmholtz coil \cite{albertazzi2013production}, was applied to the configuration. This magnetic field was oriented perpendicularly to the target surface along the $x$-axis and thus perpendicular to the main axis of the propagating plasma stream, in a manner identical to the orientation in the simulations shown in Fig 1. \textcolor{black} {Note that this configuration does not reproduce the eruptive formation stage of a CME, induced by a magnetic reconnection event \cite{Lin2015,Chen2017}, but directly the launching of the CME into the stellar magnetosphere, after it has detached from the loop at the stellar surface. } %These experimental parameters were chosen to be as close we can achieve, with the current limitations of our equipment, to be astrophysical relevant according to our scalings detailed in Table 1 and in the SI.

% \subsection{Results}

\begin{figure*}[!htp]
    \centering
    \includegraphics[width=18cm]{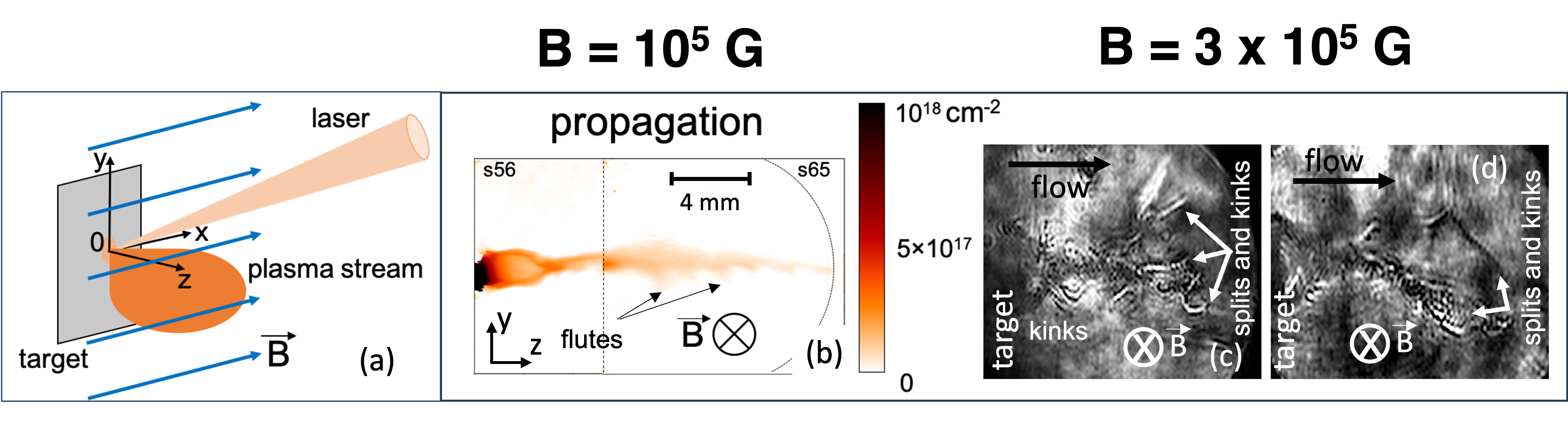}
    \caption{(a) Schematic view of the experimental set-up. A laser pulse (the light orange cone) irradiates the target surface (the grey rectangle) and creates the plasma stream (the orange oval), propagating into vacuum across the B-field lines (the blue arrows). (b-d) Two-dimensional maps of the propagating plasma stream from left to right (with the target being on the left side of the frame) in the $yz$-plane at 40 ns after the laser irradiation of the target.  (b) correspond to a magnetic field of $10^5$ G. This image shows two snapshots from two different shots with the same laser parameters, because the frame could not capture the entire plasma flow. Images in (c) and (d) correspond to a magnetic field of $3 \times 10^5$ G showing a strongly impeded propagation for two different shots (the plasma is halted in its flow at much shorter distance than what can be seen in panel (a)). The spatial scale indicated in (b) applies to all frames. In panel a is shown the line-integrated (along the x-axis) electron plasma density, coded in color (corresponding to the color bar  shown on the right of the figure). The same treatment cannot be applied to  panels b and d,  since the phase cannot be retrieved from these images, due to too strong phase shifts \textcolor{black}{(the images shown here are filtered to remove the background interference fringes. The raw interferometric images are shown in Fig.S3 of the Supp. Info. More images of flow halting in the same conditions are shown in Fig.S4 of the Supp. Info.)}.
    }
    \label{fig:Int_raw}
\end{figure*}

As already observed %The plasma itself was characterized in detail in a previous experiment (REF) and we refer the reader to the following works for the dynamics of a plasma stream propagating across a magnetic field
~\cite{Sucov1967,Bruneteau1970,Plechaty2013,garcia2016,Ivanov2017,Tang2018,PhysRevLett.123.205001,Leal2020,Filippov2021}, the laser-driven plasma is %expected to be
compressed by the magnetic field into a thin layer in the plane of the field, while it is able to extend along the direction of the magnetic field axis%~\cite{Plechaty2013,garcia2016,Ivanov2017,Tang2018,PhysRevLett.123.205001,Leal2020}
.
Our interferometry diagnostic measures two-dimensional density distributions of the plasma in the $yz$-plane and at various instants in time~\cite{HIGGINSON201748}.  \textcolor{black}{In Fig.~\ref{fig:Int_raw} we present images for two different magnetic field stengths, $10^5$ G (panel 2b) and $3 \times 10^5$ G (panel 2c-d). } For the $10^5$ G case, the plasma stream effectively propagates up to the frame edge (characterized by the circular dashed line) which corresponds to the edge of the Helmholtz coil. As characterized in detail in our previous work  \cite{PhysRevLett.123.205001}, the plasma flow is subject to the magnetized Rayleigh-Taylor instability that induces  laterally-growing flutes.
However, for $3 \times 10^5$ G magnetic field we observe a  filamentation of the plasma stream that splits in multiple branches,  with each undergoing strong kinks \textcolor{black}{(more examples of splitting and bending of halted flows recorded over different shots are shown in Fig.S4 of the Supplementary Information)}. We finally witness the halting of the flow entirely (see panel 2c and 2d for examples of two different shots), i.e. it does not propagate any further.

\begin{figure}[!ht]
\centering
    \includegraphics[width=7cm]{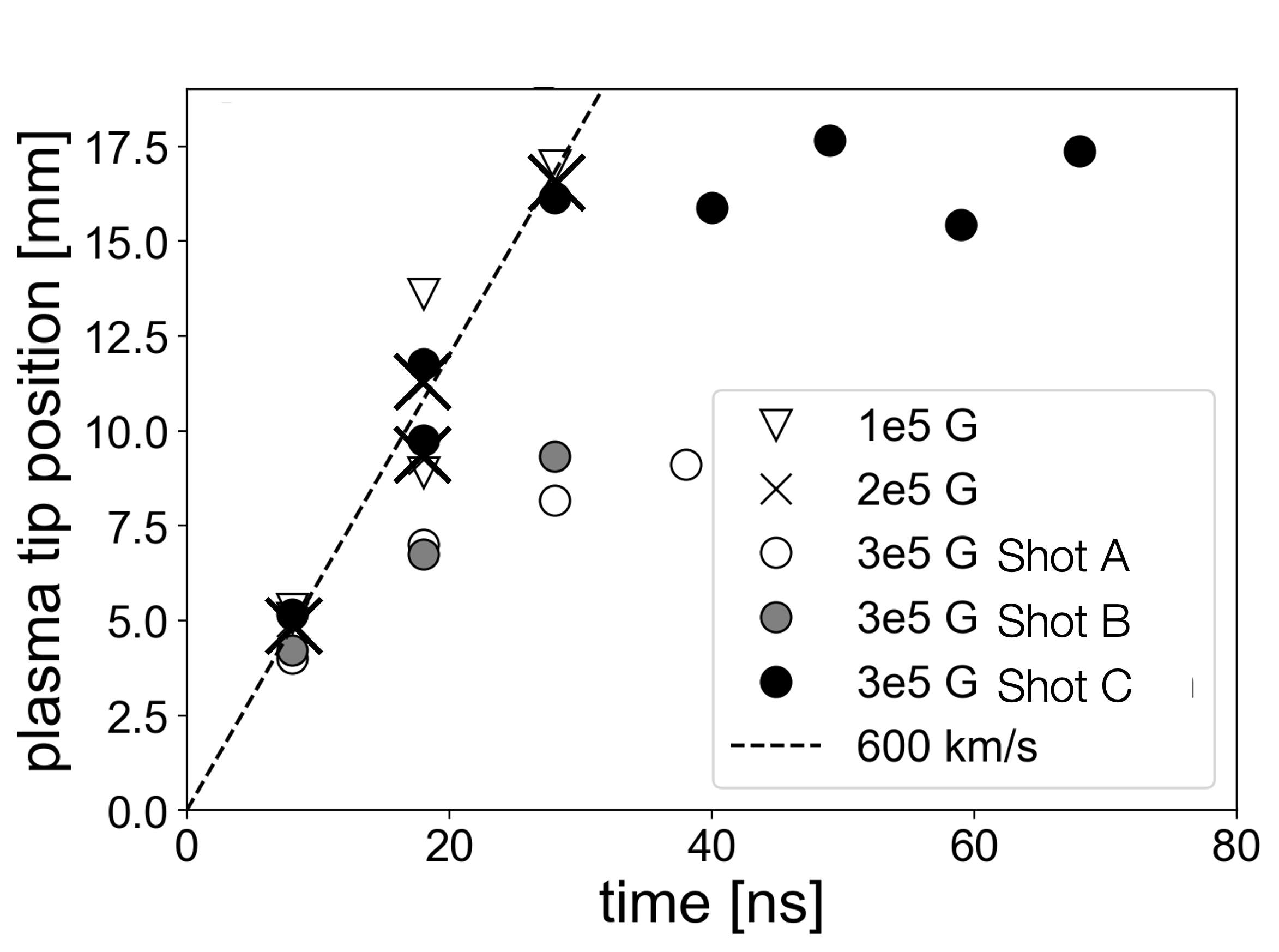}
    \caption{Temporal evolution of the tip (along the z-axis, see Fig.~\ref{fig:Int_raw}.a, with z=0 being the target surface) of the plasma flow for a range of magnetic field strengths. Various symbols correspond to the experimentally obtained data from the interferometry, the dashed lines represent constant velocity trend. \textcolor{black} {Error in the tip position is $\pm 100$ $\mu m$ given by the pixel size of the detector, i.e. smaller than the size of the symbols used in the graph.} Several shots are shown for $3 \times 10^5$ G magnetic field case to illustrate the observed variety of behaviors. Note that in the $10^5$ G and $2 \times 10^5$ G cases, we see the stream extending beyond the limit of the field of view, whereas in the case of shot C at $3 \times 10^5$ G, we see the tip stop (and oscillate, as shown) within the field of view.}
    \label{fig:Tip_speed}
\end{figure}

The temporal evolution of the plasma flow tip is shown in Fig.~\ref{fig:Tip_speed}. The tip position analysis corresponds to the density of 5$\times$10$^{16}$ cm$^{-2}$ in the two-dimensional projection of the electron density in the $yz$-plane, when the interferometry fringes distortions can still be unambiguously interpreted as being caused by the plasma.

In Fig.~\ref{fig:Tip_speed}, we observe the plasma stream propagation velocity to be around 600 km/s with no distinguishable difference between the $10^5$ G (see Fig.~\ref{fig:Int_raw}a) and $2 \times 10^5$ G magnetic fields cases. However, in the presence of a $3 \times 10^5$ G magnetic field, after the plasma stream has propagated for 20 ns, the flow stagnates. In increasing the magnetic field while keeping the plasma temperature at the target surface constant (since the laser intensity on target is kept constant),  the plasma $\beta$ lowers.

The results are therefore in excellent  consistency with the observed trend of confinement in the astrophysics simulation shown in Fig.1. We also lowered the surface plasma temperature (by lowering the laser intensity on target) and held constant the magnetic field at  $3 \times 10^5$ G like the parameter change in the astrophysical simulation. The result is qualitatively similar, although we observe a less dramatic slowing down of the plasma flow with decreasing plasma temperature (more details are presented in the Supplementary Information). This is likely due to a lower  growth rate of the instability stopping the flow, as detailed in what follows.

\section{Modeling the experimental results}
\label{sec:model}

% \subsection{Model setup}

We then performed  numerical simulations, using the  GORGON code, to understand further the origin of the observed flow disruption when changing the magnetic field strength.  GORGON is a three-dimensional resistive, single-fluid, bi-temperature and highly parallelized magneto-hydrodynamic (MHD) code \cite{chittenden2004x,ciardi2007evolution}.
Originally developed to model Z-pinches, it has been widely used to model, \textcolor{black}{and} thus benchmarked against, high-energy-density laboratory astrophysics experiments on lasers \cite{ciardi2013astrophysics,Albertazzi325,PhysRevLett.119.255002,PhysRevLett.123.205001,Filippov2021}.
In the code, the plasma is assumed to be optically thin, the ionization process is obtained from an analytical approximation to a local thermodynamic equilibrium Thomas-Fermi model, and the radiation emission is computed assuming recombination (free-bound) and Bremsstrahlung radiation \cite{salzmann1998atomic} with an upper limiter given by the black body radiation rate (Stefan-Boltzmann's law).

The simulation box is defined by a uniform Cartesian grid of dimension 8 mm $\times$ 8 mm $\times$ 24 mm and the number of cells equals $400 \times 400 \times 1200 = 1.92\times10^{8}$. The spatial resolution is homogeneous and its value is $d_x = d_y = d_z = 20\ \mu$m. The initial laser interaction with the solid target is performed using the DUED code \cite{atzeni2005fluid}, which solves the single-fluid, 3-Temperature equations in two-dimensional axisymmetric, cylindrical geometry in Lagrangian form. The laser and target parameters are taken from the experimental measurements. \textcolor{black}{The code uses the material properties of a two-temperature equation of state (EOS) model, including solid state eﬀects, and a multi-group flux-limited radiation transport module with tabulated opacities. The laser-plasma interaction is performed in the geometric optics approximation, including inverse-Bremsstrahlung absorption, which is the main laser absorption mechanism at the laser intensity used in the present experiment \cite{Atzeni2004}. }

At the end of the laser pulse (about 1 ns), the plasma profiles of density, momentum and temperature from the DUED simulations are remapped onto the three-dimensional Cartesian grid of GORGON with a superimposed magnetic field and used as initial conditions. The uniform magnetic field is aligned with the target surface (in the x-direction)\textcolor{black}{, as in the experiment,} and just as in the experiment, it has magnitudes ranging from $10^5$ G to $3 \times 10^5$ G. The purpose of the hand-off between DUED and GORGON is to take advantage of the capability of the Lagrangian code to achieve very high resolution in modeling the laser-target interaction and this simulation method has been benchmarked in a variety of similar configurations \cite{Albertazzi325,PhysRevLett.119.255002}.

\begin{figure}[!ht]
    \centering
    \includegraphics[width=0.4\textwidth]{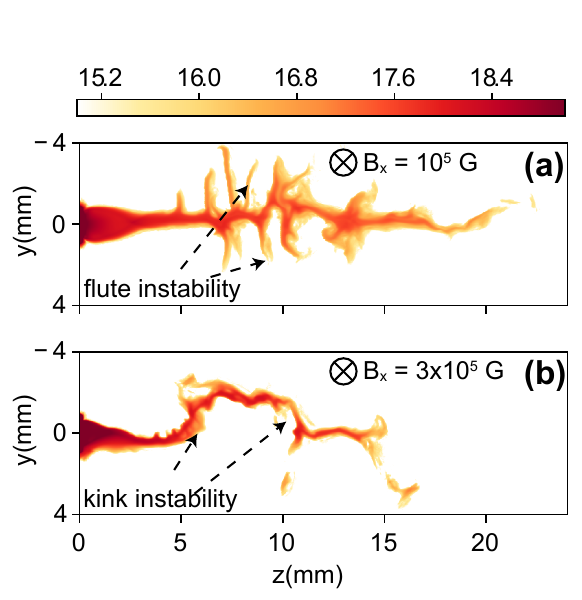}
    \caption{GORGON simulation results with integrated electron density in the $yz$-plane for a plasma flow propagating from left to right, across a magnetic field for (a) $B_x = 10^5$ G and (b) $B_x = 3 \times 10^5$ G at $t=45$ ns, with the initial velocity perturbation modes of $m=1, n=2$, as detailed in Ref. \cite{yao2022characterization}. For improved readability of the plasma, the colormap corresponds to $\log_{10} \int n_e dx$ in cm$^{-2}$, i.e., to the decimal logarithm of the electron number density integrated along the external magnetic field. The laser intensity used in the simulation is $I = 7.7 \times 10^{12}$ W/cm$^{2}$.}
    \label{fig:Bessel}
\end{figure}

% \subsection{Results}

The results of our simulations are shown in Fig.~\ref{fig:Bessel} which reports the distributions of electron density in the $yz$-plane for (a) $B_x = 10^5$ G and (b) $B_x = 3 \times 10^5$ G at $t=45$ ns. For $B_x=10^5$ G (panel a), the plasma stream propagation is stable on the whole, with flute-like instabilities growing on the sides \cite{PhysRevLett.123.205001, yao2022characterization}, consistently with the laboratory observation shown in Fig.~\ref{fig:Int_raw}b. It manages to reach $z=23$ mm at $t=45$ ns, corresponding to a velocity of 510~km/s, which is similar to what we observed experimentally in Fig.~\ref{fig:Tip_speed}.

However, when we increase the magnetic field strength from $10^5$ G to $3 \times 10^5$ G, the plasma stream propagation becomes quite unstable globally. As can be seen in Fig. \ref{fig:Bessel}b, it first bends upwards and downwards, and finally stops at $z=15$ mm, resulting in a kink-like disruption in the global morphology. This global kink-like disruption \textcolor{black}{(see Ref.\cite{Khiar2017} and references therein for detailed insights of the kinks development)} is also  in accordance with our laboratory observation shown in Fig.~\ref{fig:Int_raw}c-d, \textcolor{black}{and consistent with previous observations \cite{Hsu2005}}.
Additionally, as the magnetic field strength is increased, the plasma stream becomes thinner and the flutes structures on the sides (clearly seen with $B_x=10^5$ G) disappear. It is because the diamagnetic cavity (formed around 5 ns after the laser irradiation) collapses more quickly \cite{ciardi2013astrophysics}, leaving shorter time for the magnetized Rayleigh-Taylor instability, which is inducing these flutes, to grow \cite{yao2022characterization}.

\section{Summary}
\label{sec:conc}

By using well accepted scaling laws, it is possible to tune our laboratory plasmas to investigate physical processes as those on stars. Here we investigated one possible explanation of the lack of observational stellar CME, that they are confined by the stellar magnetic fields. We selected a plausible plasma regime of a CME that we can also produce in the laboratory. We first verified the scaled plasma regime using astrophysical simulations and we observed that, as the magnetic pressure is larger than the plasma pressure, the plasma cannot escape.

In the laboratory, we demonstrated the stopping of a laser-induced plasma stream during its propagation across a high strength externally applied magnetic field for the same plasma regime.   When the magnetic pressure strongly exceeds the plasma pressure, the propagation becomes unstable, and with the most extreme tested case, the flow stopped.

Our findings are corroborated by numerical simulations which highlight that the flow propagation disruption is likely due to the amplification of a kink instability \cite{Khiar2017} that can totally disrupt the flow. Simulations show that the plasma stream propagation becomes quite unstable globally when we increase the magnetic field strength to the values in our experiment. Such kink-like disruption in the global morphology is in excellent agreement with our optical interferometry data.

The good scalability of experimental parameters to a stellar CME thus gives weight to the hypothesis of full stopping effect of a stellar CME, propagating in the magnetic field of a star, for magnetically active stars. Therefore, our laboratory experiment can serve as an  evidence of the extreme complexity of observing stellar CMEs whose activity can be suppressed by a significant magnetic activity of the star, with important consequences for stellar mass-loss rates, angular momentum evolution, and the space weather environments of orbiting planets.

\section{Acknowledgments}\acknowledgments

The authors acknowledge the national research infrastructure Apollon and the LULI for their technical assistance. This work was supported by funding from the European Research Council (ERC) under the European Unions Horizon 2020 research and innovation program (Grant Agreement No. 787539, Project GENESIS).
This work was also supported by the Extreme Light Infrastructure Nuclear Physics (ELI-NP) Phase II, a project co-financed by the Romanian Government and the European Union through the European Regional Development Fund - the Competitiveness Operational Programme (1/07.07.2016, COP, ID 1334); the Romanian Ministry of Research and Innovation: PN23210105 (Phase 2, the Program Nucleu); and the ELI-RO grant Proiectul ELI12/16.10.2020 of the Romanian Government.  This research was also supported in part by grant NSF PHY-2309135 to the Kavli Institute for Theoretical Physics (KITP). We acknowledge GENCI-TGCC for granting us access to the supercomputer IRENE under Grants No.~A0120507594 and No.~A0130512993. The computational resources of this work were also supported by the HPC
resources of MesoPSL financed by the Region Ile de France and the project
EquipMeso (reference ANR-10-EQPX29-01) of the programme Investissements
d’Avenir supervised by the Agence Nationale pour la Recherche.

% References
%\bibliographystyle{apalike}
%\bibliographystyle{apalikenat}
%\bibliographystyle{ieeetr}
% \bibliography{report} % bibliography data in report.bib
%\bibliographystyle{spiebib} % makes bibtex use spiebib.bst
%\bibliographystyle{plainnat}
\bibliography{bib}

%\appendix

%\include{supp}

\end{document}

\title{Supplementary Information: Experimental evidence for coronal mass ejection suppression in strong stellar magnetic fields}

\author{S.N. Chen}
\affiliation{ELI NP, ”Horia Hulubei” National Institute for Physics and Nuclear Engineering, 30 Reactorului Street, RO-077125, Bucharest-Magurele, Romania}
\author{K. Burdonov}
\affiliation{LULI - CNRS, CEA, Sorbonne Université, Ecole Polytechnique, Institut Polytechnique de Paris - F-91128 Palaiseau cedex, France}
\affiliation{Sorbonne Université, Observatoire de Paris, Université PSL, CNRS, LERMA, F-75005, Paris, France}
%\affiliation{IAP, Russian Academy of Sciences, 603155, Nizhny Novgorod, Russia}
\author{W. Yao}
\affiliation{LULI - CNRS, CEA, Sorbonne Université, Ecole Polytechnique, Institut Polytechnique de Paris - F-91128 Palaiseau cedex, France}
\affiliation{Sorbonne Université, Observatoire de Paris, Université PSL, CNRS, LERMA, F-75005, Paris, France}
\author{J.~D. Alvarado-Gómez}
\affiliation{Leibniz Institute for Astrophysics Potsdam, An der Sternwarte 16, 14482 Potsdam, Germany}
\author{C. Argiroffi}
\affiliation{INAF, Osservatorio Astronomico di Palermo, Palermo, Italy}
\affiliation{University of Palermo, Department of Physics and Chemistry, Palermo, Italy}
\author{J. Béard}
\affiliation{LNCMI, UPR 3228, CNRS-UGA-UPS-INSA, 31400, Toulouse, France}
\author{S. Bolanõs}
\affiliation{LULI - CNRS, CEA, Sorbonne Université, Ecole Polytechnique, Institut Polytechnique de Paris - F-91128 Palaiseau cedex, France}
\author{R. Bonito}
\affiliation{INAF, Osservatorio Astronomico di Palermo, Palermo, Italy}
\author{A. Ciardi}
\affiliation{Sorbonne Université, Observatoire de Paris, Université PSL, CNRS, LERMA, F-75005, Paris, France}
\author{O. Cohen}
\affiliation{Lowell Center for Space Science and Technology, University of Massachusetts Lowell, 600 Suffolk St., Lowell, MA 01854, USA}
\author{J.~J.~Drake}
\affiliation{Center for Astrophysics $|$ Harvard \& Smithsonian, 60 Garden Street, Cambridge, MA 02138, USA}
\author{S. Orlando}
\affiliation{INAF, Osservatorio Astronomico di Palermo, Palermo, Italy}
\author{J. Fuchs}
\email{julien.fuchs@polytechnique.edu}
\affiliation{LULI - CNRS, CEA, Sorbonne Université, Ecole Polytechnique, Institut Polytechnique de Paris - F-91128 Palaiseau cedex, France}

\date{\today}

%\pacs{Valid PACS appear here}
% PACS, the Physics and Astronomy Classification Scheme.
\keywords{coronal mass ejections -- stellar CME -- space magnetic field}

\maketitle

\subsection{Numerical model and set-up}

AWSoM solves the three dimensional non-ideal MHD equations in conservative form for the mass, magnetic field induction, momentum, and energy. The last two equations include additional source terms resulting from the propagation and reflection of low-frequency Alfv\'en waves that drive the heating of the corona and the acceleration of the stellar wind. Sink terms such as electron heat conduction and radiative losses are also considered in the model. The reader is referred to \citet{2014ApJ...782...81V} for additional details on the model and its implementation.

In order to calculate the Alfv\'en wave turbulence spectrum, AWSoM requires information on the magnetic field within the domain. As such, the distribution of the radial magnetic field has to be provided at the inner boundary (located at $R \simeq R_{\bigstar}$), which is then used to calculate the initial state for the coronal magnetic field in the simulation via a finite difference potential field extrapolation (see \citealt{2011ApJ...732..102T}). In solar applications, the surface field information is taken from photospheric magnetic maps acquired over the course of one solar rotation (synoptic magnetograms). For the simulations analyzed here, we employ a simple $100~$G dipole aligned with the stellar rotation axis\footnote{Taken in the direction of the cartesian $z-$axis.} as inner boundary condition for $\mathbf{B}$, representing the large-scale magnetic field responsible for the CME supression. This magnetic field configuration also matches one of the scalability examples discussed in the Scaling section.

In a standard AWSoM simulation, the set of boundary conditions are complemented by specifying quantities related to the Alfv\'en wave heating mechanism. Here we take the default parameters used for solar AWSoM simulations (see \citealt{2014ApJ...782...81V}). In addition, AWSoM requires values for the density ($n_{0}$) and temperature ($T_{0}$) at the chromospheric level, in order to provide a self-consistent description of this portion of the stellar atmosphere, as well as of the transition region. This however, requires the usage of a very high-resolution grid near the inner boundary, increasing the required computation time in both, steady-state and time-dependent applications. This situation is avoided in the AWSoM-R implementation where, in the corresponding portion of the domain ($1.0~R_{\bigstar} < R \leq 1.05~R_{\bigstar}$), the plasma motion, heat transport, and Alfv\'en wave propagation, are treated in 1D following a set of magnetic field line 'threads' (see \citealt{2021ApJ...908..172S}). In this way, AWSoM-R reduces the required computation time when compared to normal AWSoM calculations (this is particularly important for large magnetic field strength values, as these significantly reduce the dynamic time-step in the simulation). The scalability test models discussed here have been therefore obtained using AWSoM-R.

The grid of our simulations is spherical, encompassing the region between $1.05~R_{\bigstar} < R \leq 30.0~R_{\bigstar}$. The angular resolution is fixed at $1.4^{\circ}$ for both meridional and azimuthal directions. The radial direction follows a $\ln(R)$ stretch profile, with the smallest grid cell reaching  $\sim$\,$0.018~R_{\bigstar}$ (close to the inner boundary), and a largest cell size of $\sim$\,$0.52~R_{\bigstar}$ (near the outer boundary). In this way, our steady-state model contains $\sim$\,6.3 million cells and a total on the order of $\sim$\,569 million in the time-dependent simulations. In terms of stellar parameters, we take solar fiducial values for mass ($M_{\star} = M_{\odot}$) and radius ($R_{\star} = R_{\odot}$), and assume a stellar rotation period\footnote{For Sun-like stars, large-scale fields on the order of $10^2$~G are expected up to this value of $P_{\rm rot}$ (see \citealt{2011IAUS..271...23D}).} of $P_{\rm rot} = 5$~d.

With the initial and boundary conditions specified, the simulation is set to evolve locally until a steady-state solution is achieved within the domain.  A basic bipolar structure is obtained, with a fast wind of rarefied plasma flowing along open field line regions, and slower and denser outflows emerging from the cusps of closed magnetic field near the equatorial plane. This corona and stellar wind configuration constitutes the starting state for a time-dependent simulation testing the scalability results discussed below.

We employ the obtained steady-state solution as the starting configuration for a time-dependent simulation. This methodology is normally used in numerical implementations of CME events in the Sun and other stars (e.g., \citealt{2017ApJ...834..173J},  \citealt{2020ApJ...895...47A,2022ApJ...928..147A}). While these studies introduce an unstable magnetic flux-rope to drive a CME in the domain, here we will follow a much simpler approach of a thermally expanding local perturbation (introduced at a certain distance from the star) representing the core of a CME, and determine whether or not is able to escape the magnetic confinement imposed by the stellar magnetic field. We consider a spherical perturbation ($R_{\rm p} = 0.15~R_{\bigstar}$) located at $R = 2R_{\bigstar}$ to match the scalability case explored in Table~\ref{tab.scal_2R}. The perturbation is introduced at rest with respect to the stellar wind frame and its evolution is followed for one hour real-time, extracting the MHD variables within the 3D domain at a cadence of one minute. We note that determining the specific influence of gravity on the scalability escape threshold goes beyond the scope of this study.

\section{Scalability}

\label{sec:scal}

In order to investigate the scalability between the laboratory experiment and the stellar CME events, we use the ``Euler similarity'' approach, developed in the works by Ryutov \cite{Ryutov_1999,Ryutov_2000,doi:10.1063/1.5042254}.

The general idea of the approach is as follows. Two systems should evolve identically when the following two scaling quantities are held constant: the Euler number $E_u = V(\rho/p)^{1/2}$ (here and below the formulas are presented for cgs units), and the thermal plasma beta $\beta_{T}  = 8\pi p  / B^{2}$ (B is the magnetic field, $p = k_{B}(n_i T_{i} + n_{e} T_{e})$ is the thermal pressure, $k_{B}$ is the Boltzmann constant and $n_{i,e}$ and $T_{i,e}$ are the number densities and temperatures of the ions and electrons, respectively). To take into account the ram pressure of the flow, we also introduce the modified plasma beta $\beta_{mod} = \beta_{T} + \beta_{dyn} = (8\pi p + \rho V^{2}) / B^{2}$, where V is the flow velocity and $\rho$ is the mass density.

To exclude the impact of dissipative processes affecting the fluid dynamics, and to use ideal MHD equations, several following parameters should be higher than 1 for both systems.
The first is the Reynolds number, $R_e = LV/\nu$ (the ratio of the inertial force to the viscous force), which is responsible for the viscous dissipation, where $L$ is the characteristic spatial scale, $V$ is the flow velocity, and
\begin{equation}
\nu [cm^2 s^{-1}] = 1.4 \times 10^9 \frac{T_i [kK]^{2.5}}{\Lambda \sqrt{A} Z^4 n_i}
\end{equation}
is the kinematic viscosity, where $\Lambda$ is the Coulomb logarithm and A is the averaged atomic weight (\cite{Ryutov_1999} p. 825).

The second is the magnetic Reynolds number, $R_{eM} = LV/\eta$ (the ratio of the convection over Ohmic dissipation), which is responsible for the resistive diffusion. To derive $R_{eM}$, we need to evaluate the magnetic diffusivity,
% \begin{equation}
% \eta = \frac {c^2}{4 \pi \sigma} = \frac{m_e c^2}{4 \pi n_e e^2 \tau_{col\:e}}
% ,\end{equation}
% where $\sigma$ is the electrical conductivity, $m_e$ is the electron mass, $e$ is the electron charge, and
% % $c^2 = 1$
% $\tau_{col\:e}$ is the electron collisional time
% % (\cite{Ryutov_2000} p. 467).
% \begin{equation}
% \eta [cm^2 s^{-1}] = 5.0 \times 10^{8} \frac{\Lambda Z}{T_e [kK]^{1.5}}.
% \end{equation}
\begin{equation}
\eta [cm^2 s^{-1}] = 9.8 \times 10^{7} \frac{\Lambda Z}{T_e [kK]^{1.5}}.
\end{equation}
(\cite{Ryutov_2000} p. 467).

And the third is the Peclet number $P_e = LV/\chi$ (the ratio of heat convection to the heat conduction). This last parameter is the one responsible for the thermal conduction. It involves the thermal diffusivity (\cite{Ryutov_1999} p. 824):
\begin{equation}
\chi [cm^2 s^{-1}] = 1.4 \times 10^{11} \frac{T_i [kK]^{2.5}}{\Lambda Z (Z+1) n_i}.
\end{equation}

%the Reynolds number (the ratio of the inertial force to the viscous force), which is responsible for the viscous dissipation, the magnetic Reynolds number (the ratio of the convection over Ohmic dissipation), responsible for the resistive diffusion, and the Peclet number (the ratio of heat convection to the heat conduction), responsible for the thermal conduction, should be higher than 1 for both systems of interest.

Because of the scarcity of observational data, there is little information about typical CME parameters \cite{Argiroffi2019}.
%{\bf{\color{red}see end of the section}}
Some estimations can be found in \cite{10.1093/mnras/stx1969,10.1093/mnras/staa1021}, supposing CME velocities from hundreds to thousands of kilometers per second, and the effective star surface temperatures of the order of kilo-Kelvins. However, as can be seen from \cite{Argiroffi2019}, the temperature of the CME itself can be up to tens of mega-Kelvins.

In the MHD modelling presented in \cite{Alvarado_G_mez_2018}, it was shown that a magnetic field of 75 G is able to fully confine a CME that is analogous to a typical solar CME. And in \cite{Alvarado_G_mez_2019} the impact of much stronger magnetic fields up to 1400 G was investigated. The parameters mentioned in that article are not scalable to our laboratory conditions, but the same time, there is also no observational proofs of the existence of stellar CMEs with such parameters.

Finally, looking at Table 1 of the Supplementary Information, we can see that there is a good scaling between the simulated stellar CME and the laboratory flow, since the three quantities, the Euler number, the thermal beta and the modified beta, are all in close proximity between the two systems. Further, still based on Ryutov's analysis, we can extract the scaling factors in space, density, and velocity between the laboratory plasma flow and the CME.  For magnetic field of $3 \times 10^5$ G and for the nominal laser irradiance of $10^{13}$ W/cm$^{2}$ versus a CME with temperature 230 kK propagating across 10 G magnetic field, the spatial scaling parameters are as follows:
\begin{equation}
a = \frac{r_{ast}}{r_{lab}} = \frac{5 \times 10^{7} {\rm [km]}}{1 {\rm [cm]}} = 5 \times 10^{11},
\end{equation}
\noindent
The density scaling parameters are
\begin{equation}
b = \frac{\rho_{ast}}{\rho_{lab}} = \frac{2.1 \times 10^{-14} [{\rm g /cm}^{3}]}{3.7 \times 10^{-5} [{\rm g / cm}^{3}]} = 5.7 \times 10^{-10},\
\end{equation}
\noindent
and the velocity scaling parameters are written as:
\begin{equation}
c = \frac{V_{ast}}{V_{lab}} = \frac{800 [{\rm km s}^{-1}]}{600 [{\rm km s}^{-1}]} = 1.3~,
\end{equation}
\noindent
Thus, the temporal scaling $t_{ast}$ = $(a / c )t_{lab}$ gives that 10 ns in the laboratory case with $3 \times 10^5$ G magnetic field is equivalent to around $3.8 \times 10^{12}$ ns ($\sim$ 1 hour) for stellar CME propagating through $B_{ast}$ = $B_{lab} c \sqrt{b} = 10$ G magnetic field at
2
% 5
stellar radii.

\begin{table*}
\centering{}%
\begin{tabular}{cc||c|c|c}
 & \multicolumn{1}{c||}{\textbf{\textit{Laboratory}}} & \multicolumn{1}{c}{\textbf{\textit{}}}  & \multicolumn{1}{c}{\textbf{\textit{stellar CME}}} & \multicolumn{1}{c}{\textbf{\textit{}}}\tabularnewline

\hline
\hline
material & \multicolumn{1}{c||}{$CF_{2}$ (teflon)} & \multicolumn{1}{c}{} & \multicolumn{1}{c}{$H$ } & \multicolumn{1}{c}{}\tabularnewline

charge state Z & \multicolumn{1}{c||}{$\textbf{5.9}$} & \multicolumn{1}{c}{} & \multicolumn{1}{c}{$\textbf{1}$} & \multicolumn{1}{c}{}\tabularnewline

mass number A & \multicolumn{1}{c||}{$\textbf{17.3}$} & \multicolumn{1}{c}{} & \multicolumn{1}{c}{$\textbf{1.28}$} & \multicolumn{1}{c}{}\tabularnewline

spatial scale $[cm]$ & \multicolumn{1}{c||}{$\mathbf{1}$} & \multicolumn{1}{c}{} & \multicolumn{1}{c}{$\mathbf{2 \times 10^{11}}$} & \multicolumn{1}{c}{}\tabularnewline

%spatial scale $[cm]$ & $\mathbf{0.1}$ & $\mathbf{}$ & $\mathbf{5 \times 10^{11}}$ & $\mathbf{7 \times 10^{11}}$\tabularnewline

B-field $[G]$ & \multicolumn{1}{c||}{$\mathbf{3 \times 10^{5}}$} & \multicolumn{1}{c}{} & \multicolumn{1}{c}{$\mathbf{100}$} & \multicolumn{1}{c}{}\tabularnewline

electron density $[cm^{-3}]$ & \multicolumn{1}{c||}{ $\mathbf{7.5 \times10^{18}}$} & \multicolumn{1}{c}{} & \multicolumn{1}{c}{$\mathbf{10^{11}}$} & \multicolumn{1}{c}{} \tabularnewline

mass density $[g.cm^{-3}]$ &  \multicolumn{1}{c||}{$3.7 \times10^{-5}$} & \multicolumn{1}{c}{} & \multicolumn{1}{c}{$2.1 \times10^{-13}$} & \multicolumn{1}{c}{}\tabularnewline

electron temperature $[kK]$ &  \multicolumn{1}{c||}{$\mathbf{500}$} & \multicolumn{1}{c}{} & \multicolumn{1}{c}{$\mathbf{3.5\times 10^3}$} & \multicolumn{1}{c}{}\tabularnewline

ion temperature $[kK]$ & \multicolumn{1}{c||}{ $\mathbf{500}$} & \multicolumn{1}{c}{} & \multicolumn{1}{c}{$\mathbf{3.5 \times 10^3}$} & \multicolumn{1}{c}{}\tabularnewline

flow velocity $[km.s^{-1}]$ & \multicolumn{1}{c||}{ $\mathbf{600}$} & \multicolumn{1}{c}{} & \multicolumn{1}{c}{$\mathbf{4.6 \times 10^3}$} & \multicolumn{1}{c}{}\tabularnewline

sound velocity $[km.s^{-1}]$ & \multicolumn{1}{c||}{ $51$} & \multicolumn{1}{c}{} & \multicolumn{1}{c}{$270$} & \multicolumn{1}{c}{}\tabularnewline

Alfv\'en velocity $[km.s^{-1}]$ & \multicolumn{1}{c||}{ $140$} & \multicolumn{1}{c}{} & \multicolumn{1}{c}{$610$} & \multicolumn{1}{c}{}\tabularnewline

Coulomb logarithm & \multicolumn{1}{c||}{ $7.3$ }& \multicolumn{1}{c}{} & \multicolumn{1}{c}{$18.4$}  & \multicolumn{1}{c}{}\tabularnewline

electron mean free path $[cm]$ & \multicolumn{1}{c||}{ $7.3 \times10^{-5}$ }& \multicolumn{1}{c}{} & \multicolumn{1}{c}{$7.1 \times 10^{5}$}  & \multicolumn{1}{c}{}\tabularnewline

electron collision time $[ns]$ & \multicolumn{1}{c||}{ $2.8 \times10^{-4}$ }& \multicolumn{1}{c}{} & \multicolumn{1}{c}{$9.8 \times 10^{5}$} & \multicolumn{1}{c}{}\tabularnewline

electron Larmor radius $[cm]$ & \multicolumn{1}{c||}{ $5 \times10^{-5}$} & \multicolumn{1}{c}{} & \multicolumn{1}{c}{$0.4$}& \multicolumn{1}{c}{}\tabularnewline

electron gyrofrequency $[s^{-1}]$ & \multicolumn{1}{c||}{ $8.4 \times10^{11}$} & \multicolumn{1}{c}{} & \multicolumn{1}{c}{$2.8 \times 10^{8}$}  & \multicolumn{1}{c}{}\tabularnewline

ion mean free path $[cm]$ & \multicolumn{1}{c||}{ $2.9 \times10^{-6}$ } & \multicolumn{1}{c}{} & \multicolumn{1}{c}{$3.1 \times 10^{7}$} & \multicolumn{1}{c}{}\tabularnewline

ion collision time $[ns]$ & \multicolumn{1}{c||}{ $2 \times10^{-3}$} & \multicolumn{1}{c}{} & \multicolumn{1}{c}{$6.7 \times 10^{7}$} & \multicolumn{1}{c}{}\tabularnewline

ion Larmor radius $[cm]$ & \multicolumn{1}{c||}{ $1.5 \times10^{-3}$ }& \multicolumn{1}{c}{} & \multicolumn{1}{c}{$6.2$}  & \multicolumn{1}{c}{}\tabularnewline

ion gyrofrequency $[s^{-1}]$ & \multicolumn{1}{c||}{ $1.6 \times10^{8}$} & \multicolumn{1}{c}{} & \multicolumn{1}{c}{$1.2 \times10^{5}$} & \multicolumn{1}{c}{}\tabularnewline

Mach number $M$ & \multicolumn{1}{c||}{ $11.9$} & \multicolumn{1}{c}{} & \multicolumn{1}{c}{$16.8$}& \multicolumn{1}{c}{}\tabularnewline

%Alfven Mach number $M_{alf}$ & \multicolumn{1}{c||}{ $4.3$} & \multicolumn{1}{c}{} & \multicolumn{1}{c}{$4.1$} & \multicolumn{1}{c}{}\tabularnewline

% magnetic diffusion time $[ns]$ & \multicolumn{1}{c||}{ $7.3 \times 10^{3}$ }& \multicolumn{1}{c}{} & \multicolumn{1}{c}{$7.7 \times 10^{27}$}& \multicolumn{1}{c}{} \tabularnewline

Alfven Mach number $M_{Alf}$  & \multicolumn{1}{c||}{4.3} & \multicolumn{1}{c}{} & \multicolumn{1}{c}{7.6} & \multicolumn{1}{c}{}\tabularnewline

Reynolds number $Re$ & \multicolumn{1}{c||}{$1.3 \times10^{4}$ } & \multicolumn{1}{c}{} & \multicolumn{1}{c}{$ 1.5 \times 10^{11}$} & \multicolumn{1}{c}{} \tabularnewline

Magnetic Reynolds number $R_{eM}$ & \multicolumn{1}{c||}{$5 \times 10^{4}$} & \multicolumn{1}{c}{} & \multicolumn{1}{c}{$3.2 \times 10^{18}$} & \multicolumn{1}{c}{}\tabularnewline

Peclet number $P_e$ & \multicolumn{1}{c||}{$74$} & \multicolumn{1}{c}{} & \multicolumn{1}{c}{$3.1\times 10^{9}$} & \multicolumn{1}{c}{}\tabularnewline

Euler number $Eu$ & \multicolumn{1}{c||}{$15$} & \multicolumn{1}{c}{} & \multicolumn{1}{c}{$22$} & \multicolumn{1}{c}{}\tabularnewline

thermal plasma beta $\beta_T$ & \multicolumn{1}{c||}{$0.17$} & \multicolumn{1}{c}{} & \multicolumn{1}{c}{0.24} & \multicolumn{1}{c}{}  \tabularnewline

modified plasma beta $\beta_{mod}$ & \multicolumn{1}{c||}{$37$}  & \multicolumn{1}{c}{} &\multicolumn{1}{c}{11}  & \multicolumn{1}{c}{} \tabularnewline

\tabularnewline
\end{tabular}
\caption{Comparison and scalability between the laboratory plasma stream, created by the high laser irradiance, propagating across the magnetic field of 3$\times10^5$ G, and a hypothetical stellar CME event, propagating across the magnetic field of 100 G at the distance of 2 stellar radii. The primary parameters are in bold, while the numbers in light are derived from these primary numbers.}
\label{tab.scal_2R}
\end{table*}

\clearpage

\section{Plasma flow at lower temperature}

In the astrophysical simulation performed with AWSoM-R, to observe changes in the plasma confinement evolution, the initial plasma temperature was decreased to lower the plasma $\beta$.  We had performed a similar plasma temperature reduction in the laboratory by decreasing the laser intensity to $10^{12}$ W/cm$^{2}$ on target. For the comparisons below, we use the flow rate as it is easily determined by our interferometry diagnostic.

\begin{figure}[!ht]
    \centering
    \includegraphics[width=7cm]{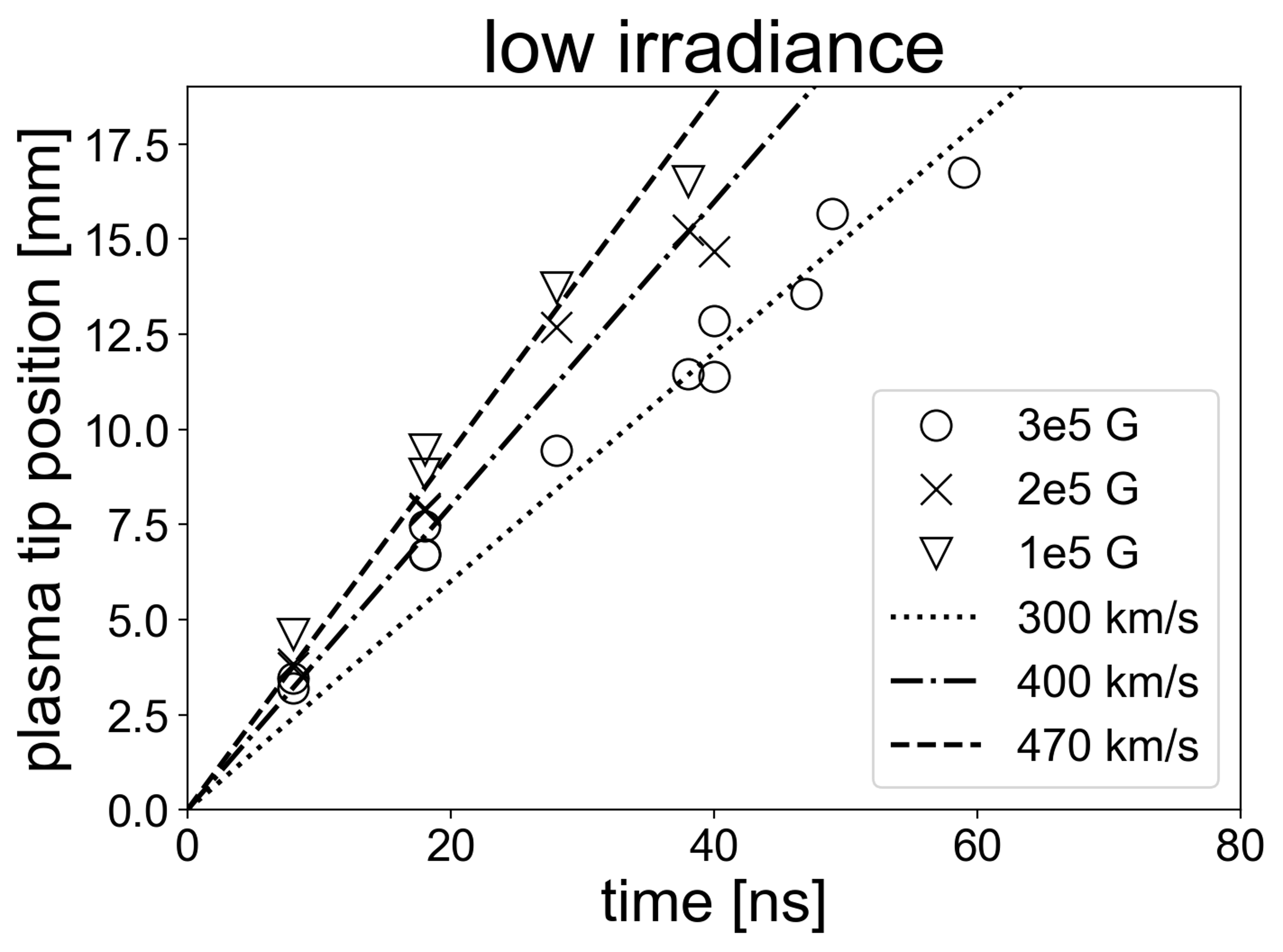}

    \caption{Temporal evolution of the tip of the plasma flow for the low laser irradiance case. Various symbols correspond to the experimentally obtained data from the interferometry, dashed, dotted and dashed-dotted lines represent constant velocity trends.}
    \label{fig:Tip_speed}
\end{figure}

At this lower laser irradiance, we can first see a particular trend when the magnetic field is varied. As shown in Figure~\ref{fig:Tip_speed}, we show the position of the tip of the flow in time. We see that with the increase in magnetic field strength, there is a slowing down just as in the case of higher laser intensity, but the effect is less abrupt. For magnetic field strengths of $10^5$ G and $2 \times 10^5$ G respectively, the velocity remains constant 470~km~s$^{-1}$ and 400~km~s$^{-1}$ respectively. For the $3 \times 10^5$ G case, the plasma stream undergoes a reduction of propagation velocity as it starts out at 400 km/s and decreases to 300 km/s around t=30 ns.

\begin{figure}[!ht]
    \centering
    \includegraphics[width=7cm]{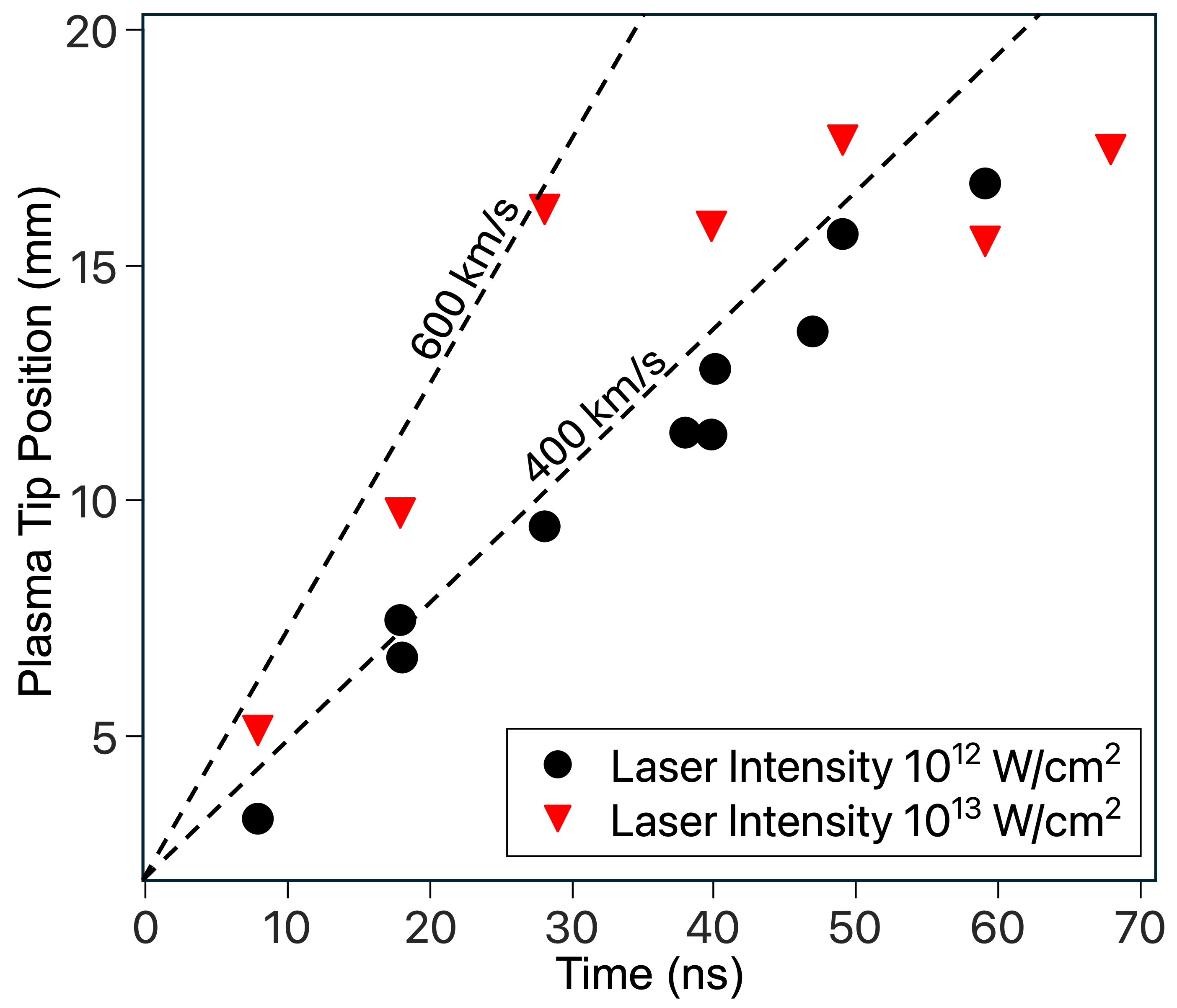}

    \caption{Comparison of the evolution of the tip of the plasma flow between low and high laser irradiance. Various symbols correspond to the experimentally obtained data from the interferometry, dashed, dotted and dashed-dotted lines represent constant velocity trends.}
    \label{fig:Tip_speed_vs}
\end{figure}

In Figure ~\ref{fig:Tip_speed_vs}, we compare the tip position in time for the two laser irradiance levels while keeping the magnetic field at  $3 \times 10^5$ G. As evidenced by the comparisons made in  Figure ~\ref{fig:Tip_speed} in the Supplementary Information and in Figure 3 of the main article, the flow is slowing down in both cases, but in a much more abrupt manner in the case of the higher (nominal) laser irradiance.

\section{Supplementary Experimental Data}

\label{sec:expt}

\begin{figure*}[!ht]
    \centering
    \includegraphics[width=16cm]
    {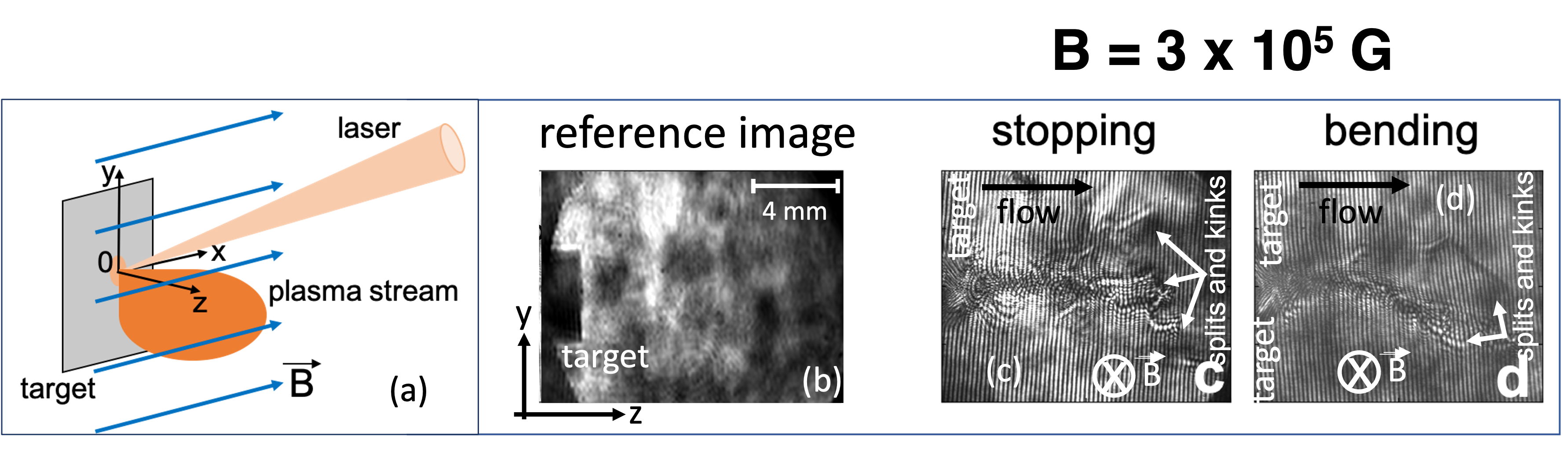}

    \caption{ \textcolor{black}{(a) Schematic view of the experimental set-up. A laser pulse (the light orange cone) irradiates the target surface (the grey rectangle) and creates the plasma stream (the orange oval), propagating into vacuum across the B-field lines (the blue arrows). (b-d) Two-dimensional maps of the propagating plasma stream from left to right (with the target being on the left side of the frame) in the $yz$-plane at 40 ns after the laser irradiation of the target. In panel a is shown the line-integrated (along the x-axis) electron plasma density, coded in color (corresponding to the color bar  shown on the right of the figure). In panels b and d, we show the raw interferograms since the phase cannot be retrieved from these images, due to too strong phase shifts. (b) correspond to a magnetic field of $10^5$ G. This image shows two snapshots from two different shots with the same laser parameters, because the frame could not capture the entire plasma flow. Images in (c) and (d) correspond to a magnetic field of $3 \times 10^5$ G showing a strongly impeded propagation for two different shots. The spatial scale indicated in (b) applies to all frames.}}
    \label{fig:SI_expt}
\end{figure*}

\textcolor{black}{In Fig.~\ref{fig:SI_expt} we present the raw interferograms that correspond to Fig 2 of the main manuscript.  The first panel (a) presents the cartoon of the experimental setup and (b) shows the location of the target prior to laser irradiation, illustrating the typical light intensity distribution of the probing laser beam. Panels (c) and (d) correspond to Fig 2(c) and Fig 2(d) of the main manuscript respectively. Despite the inhomogeneous light intensity distribution of the probe beam (illustrated in panel(a)), one can clearly see the expanding plasma, being subject to breakup and filamentation.}

\begin{figure*}[!ht]
    \centering
    \includegraphics[width=16cm]
    {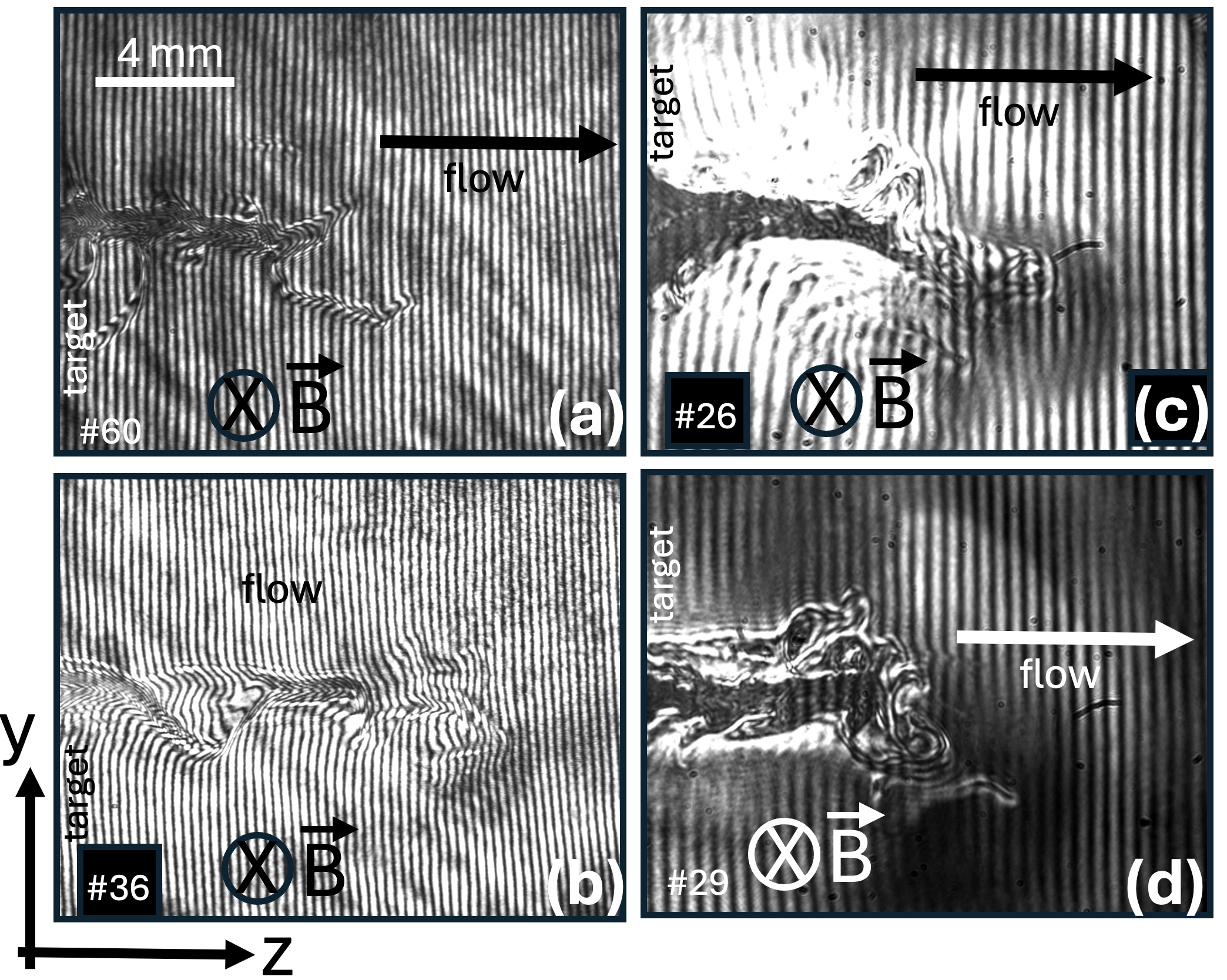}

    \caption{ \textcolor{black}{(a-d) Two-dimensional raw interferograms of the propagating plasma stream from left to right (with the target being on the left side of the frame) in the $yz$-plane at 40 ns after the laser irradiation of the target and for different shots (with the same laser parameters in the high-irradiance case).  All shots correspond to a magnetic field of $3 \times 10^5$ G. They illustrate different behavior of the plasma, all displaying its halting, filamentation or bending. This is contrary to the situation (illustrated in Fig.2.b of the main text) where the magnetic field is lowered to  $10^5$ G, in which case the plasma propagates unimpeded.  The spatial scale indicated in (a) applies to all frames.}}
    \label{fig:SI_expt2}
\end{figure*}

\textcolor{black}{In Fig.~\ref{fig:SI_expt2} we present four other shots  that correspond to the same situation as of Fig.~\ref{fig:SI_expt}, i.e. of the plasma being halted by the $3 \times 10^5$ G magnetic field, illustrating different types of flow break-up or bending. }

\clearpage

\bibliography{bib}

%\appendix

%\include{supp}